\documentclass[useAMS,usenatbib]{mn2e}
\usepackage{natbib}
\usepackage{threeparttable}
\usepackage{longtable}
\usepackage{environ}
\usepackage{graphicx}
\usepackage{textcomp}
\usepackage[fleqn]{amsmath}
\usepackage{lscape}
\usepackage{pdflscape}
\usepackage{dcolumn}
\usepackage{rccol}
\usepackage{threeparttablex}
\usepackage{amssymb}
\usepackage{comment} 
\usepackage{gensymb}
\usepackage{lineno}
\usepackage{eso-pic}

\AddToShipoutPictureBG*{%
  \AtPageUpperLeft{%
    \hspace{0.75\paperwidth}%
    \raisebox{-3.5\baselineskip}{%
      \makebox[0pt][l]{\textnormal{DES 2016-0213}}
}}}%

\AddToShipoutPictureBG*{%
  \AtPageUpperLeft{%
    \hspace{0.75\paperwidth}%
    \raisebox{-4.5\baselineskip}{%
      \makebox[0pt][l]{\textnormal{FERMILAB-PUB-16-614-AE}}
}}}%

\newcommand{\about}{$\sim\!\!$~}

\newcommand{\mstellar}{\ensuremath{M_{\mathrm{stellar}}}}
\newcommand{\omatter}{\ensuremath{\Omega_{\mathrm{M}}}}

\newcommand{\ciii}{\ensuremath{\mathrm{C}\,\textsc{iii}}}

\newcommand{\mgi}{\ensuremath{\mathrm{Mg}\,\textsc{i}}}
\newcommand{\mgii}{\ensuremath{\mathrm{Mg}\,\textsc{ii}}}
\newcommand{\Feii}{\ensuremath{\mathrm{Fe}\,\textsc{ii}}}
\newcommand{\Feiii}{\ensuremath{\mathrm{Fe}\,\textsc{iii}}}

\newcommand{\Cofs}{$^{56}$Co}

\title[DES15E2mlf]{DES15E2mlf: A Spectroscopically Confirmed Superluminous Supernova that Exploded 3.5\,Gyr After the Big Bang}
\author[Pan et al.]{
Y.-C.~Pan$^{1}$\thanks{E-mail:ypan6@ucsc.edu},
R.~J.~Foley$^{1}$,
M.~Smith$^{2}$,
L.~Galbany$^{3,4}$,
C.~B.~D'Andrea$^{2,5}$,\newauthor
S.~Gonz{\'a}lez-Gait{\'a}n$^{6,7}$,
M.~J.~Jarvis$^{8,9}$,
R.~Kessler$^{10,11}$,
E.~Kovacs$^{12}$,
C.~Lidman$^{13}$,\newauthor
R.~C.~Nichol$^{5}$,
A.~Papadopoulos$^{5,14}$,
M.~Sako$^{15}$,
M.~Sullivan$^{2}$,
T.~M.~C.~Abbott$^{16}$,\newauthor
F.~B.~Abdalla$^{17,18}$,
J.~Annis$^{19}$,
K.~Bechtol$^{20}$,
A.~Benoit-L{\'e}vy$^{17,21,22}$,
D.~Brooks$^{17}$,\newauthor
E.~Buckley-Geer$^{19}$,
D.~L.~Burke$^{23,24}$,
A.~Carnero~Rosell$^{25,26}$,
M.~Carrasco~Kind$^{27,28}$,\newauthor
J.~Carretero$^{29,30}$,
F.~J.~Castander$^{29}$,
C.~E.~Cunha$^{23}$,
L.~N.~da Costa$^{25,26}$,
S.~Desai$^{31}$,\newauthor
H.~T.~Diehl$^{19}$,
P.~Doel$^{17}$,
T.~F.~Eifler$^{32}$,
D.~A.~Finley$^{19}$,
B.~Flaugher$^{19}$,\newauthor
J.~Frieman$^{10,19}$,
J.~Garc\'ia-Bellido$^{33}$,
D.~A.~Goldstein$^{34,35}$,
D.~Gruen$^{23,24}$,\newauthor
R.~A.~Gruendl$^{27,28}$,
J.~Gschwend$^{25,26}$,
G.~Gutierrez$^{19}$,
D.~J.~James$^{16,36}$,
A.~G.~Kim$^{35}$,\newauthor
E.~Krause$^{23}$,
K.~Kuehn$^{13}$,
N.~Kuropatkin$^{19}$,
O.~Lahav$^{17}$,
M.~Lima$^{25,37}$,\newauthor
M.~A.~G.~Maia$^{25,26}$,
M.~March$^{15}$,
J.~L.~Marshall$^{38}$,
P.~Martini$^{39,40}$,
R.~Miquel$^{30,41}$,\newauthor
P.~Nugent$^{34,35}$,
A.~A.~Plazas$^{32}$,
A.~K.~Romer$^{42}$,
E.~Sanchez$^{43}$,
V.~Scarpine$^{19}$,\newauthor
M.~Schubnell$^{44}$,
I.~Sevilla-Noarbe$^{43}$,
R.~C.~Smith$^{16}$,
F.~Sobreira$^{25,45}$,\newauthor
E.~Suchyta$^{46}$,
M.~E.~C.~Swanson$^{28}$,
R.~C.~Thomas$^{35}$,
A.~R.~Walker$^{16}$\\
\\  \LARGE\textup{\centerline{(The DES Collaboration)}}
}

\begin{document}
\maketitle

\label{firstpage}
\begin{abstract}
  We present the Dark Energy Survey (DES) discovery of DES15E2mlf, the most
  distant superluminous supernova (SLSN) spectroscopically confirmed to date. The light curves 
  and Gemini spectroscopy of DES15E2mlf indicate that it is a Type I 
  superluminous supernova (SLSN-I) at $z = 1.861$ (a lookback time of $\sim10$\,Gyr) 
  and peaking at $M_{\rm AB} = -22.3\pm0.1$\,mag.
  Given the high redshift, our data probe the rest-frame
  ultraviolet (1400--3500\,\AA) properties of the SN, finding velocity
  of the \ciii\ feature changes by $\sim5600$\,km\,$\rm s^{-1}$ over 14~days
  around maximum light.  We find the host galaxy of DES15E2mlf
  has a stellar mass of $3.5^{+3.6}_{-2.4} \times 10^{9}$~M$_{\sun}$, which is
  more massive than the typical SLSN-I host galaxy.
\end{abstract}


\section{Introduction}
\label{sec:introduction}

In the era of untargeted wide-field sky surveys, thousands of
supernovae (SNe) are discovered each year, uncovering many new classes
of SNe. Among these, superluminous supernovae
\citep[SLSNe;][]{2012Sci...337..927G}, being 10--100 times as bright
as typical Type Ia SNe (SNe Ia) and core-collapse SNe (CCSNe), are
particularly interesting.  SLSNe are relatively rare, having a rate
$<$0.1\% that of the CCSN rate \citep[e.g.,][]{2013MNRAS.431..912Q,
  2015MNRAS.448.1206M,2016arXiv160505250P}.  Given their high luminosities, SLSNe are
detectable to higher redshifts compared with SNe Ia and CCSNe \citep[e.g.,
$z\gtrsim1.5$;][]{2012ApJ...755L..29B, 2013ApJ...779...98H,2012Natur.491..228C}.

There is significant debate about what powers SLSNe.  Several models
have been proposed to explain the extremely high luminosities of these
events.  For instance, the SN luminosity could be primarily powered by
the spin down of a newborn magnetar, where the rotation energy is
deposited into the SN ejecta \citep{2010ApJ...717..245K,
  2010ApJ...719L.204W}.  Alternatively, the release of shock energy by
collision with an opaque circumstellar material
\cite[CSM;][]{2011ApJ...729L...6C, 2012ApJ...757..178G}, or the
explosion of pair-instability supernova
\citep[PISN;][]{1967PhRvL..18..379B} could power the high luminosity.

\citet{2012Sci...337..927G} divided SLSNe into three different
sub-classes: SLSN-I, SLSN-II and SLSN-R, based on either their
spectroscopic or photometric properties.  SLSNe-II have strong, narrow
hydrogen emission lines in their spectra, which are believed to be
caused by the interaction with hydrogen-rich CSM
\citep[e.g.,][]{2007ApJ...666.1116S}.  SLSNe-R have a slow
luminosity decay, consistent with being powered by \Cofs\ decay, a
predicted outcome of pair-instability models
\citep[e.g.,][]{2009Natur.462..624G}.

Compared to SLSNe-II, SLSNe-I are characterized by the lack of
hydrogen features in their spectra and exceptionally high peak
luminosities \citep{2011Natur.474..487Q}. The spectra at early phases
are usually featureless in the optical region with most of the flux
and absorption features found in the ultraviolet \citep[UV,
e.g.,][]{2011Natur.474..487Q, 2011ApJ...743..114C,
  2013ApJ...779...98H, 2014ApJ...797...24V, 2015MNRAS.449.1215P,
  2016ApJ...818L...8S}.  These events show some spectral similarities
with broad-line type Ic SNe, but require an additional energy source
beyond that of standard CCSNe \citep[e.g., ][]{2013ApJ...770..128I,
  2013Natur.502..346N}.

Previous studies have found that SLSNe-I exclusively reside in faint,
low-mass ($M_{\rm stellar} < 10^{10} M_{\sun}$) galaxies
\citep{2011ApJ...727...15N, 2013ApJ...763L..28C, 2013ApJ...771...97L,
  2014ApJ...787..138L, 2015MNRAS.452.1567C, 2015MNRAS.449..917L}.
For the subset of host galaxies having spectroscopic metallicity measurements,
all have low gas-phase metallicities.  Similarly, SLSN host
galaxies have high specific star-formation rates (sSFR), making them
similar to the host galaxies of long-duration gamma-ray bursts
\citep[although it is possible that the SLSN progenitor star populations 
are even {\it younger}]{2015MNRAS.449..917L}.  The low metallicity environments
have been interpreted as being critical for the formation of SLSNe and
are used as a distinguishing characteristic for models \citep[e.g.,][]{2016arXiv160408207P}.

The Dark Energy Survey supernova program \citep[DES-SN;][]{2012ApJ...753..152B} has discovered
$\sim20$ SLSNe and identified many additional SLSNe candidates since 2013 
\citep{2015MNRAS.449.1215P,2016ApJ...818L...8S,2016MNRAS.460.1270D}.
Here we present DES15E2mlf, another interesting SLSN at $z = 1.861$
discovered by the DES-SN.  DES15E2mlf is the highest-redshift spectroscopically
confirmed SLSN, providing unique insight into the UV properties and
possible redshift evolution of SLSNe.  A plan of the paper follows.
In Section~\ref{sec:obs} we describe the discovery and observations of
DES15E2mlf. Section~\ref{sec:analysis} presents the analyses of the photometry,
spectroscopy, and host galaxy.  We discuss our findings in
Section~\ref{sec:discussion} and conclude in
Section~\ref{sec:conclusion}.  Throughout this paper, we assume
$\mathrm{H}_{0} = 70$\,km\,s$^{-1}$\,Mpc$^{-1}$ and a flat universe
with $\omatter = 0.3$.

\section{Discovery and Observations}
\label{sec:obs}
\begin{figure}
	\centering
	\begin{tabular}{c}
		\includegraphics*[scale=0.51]{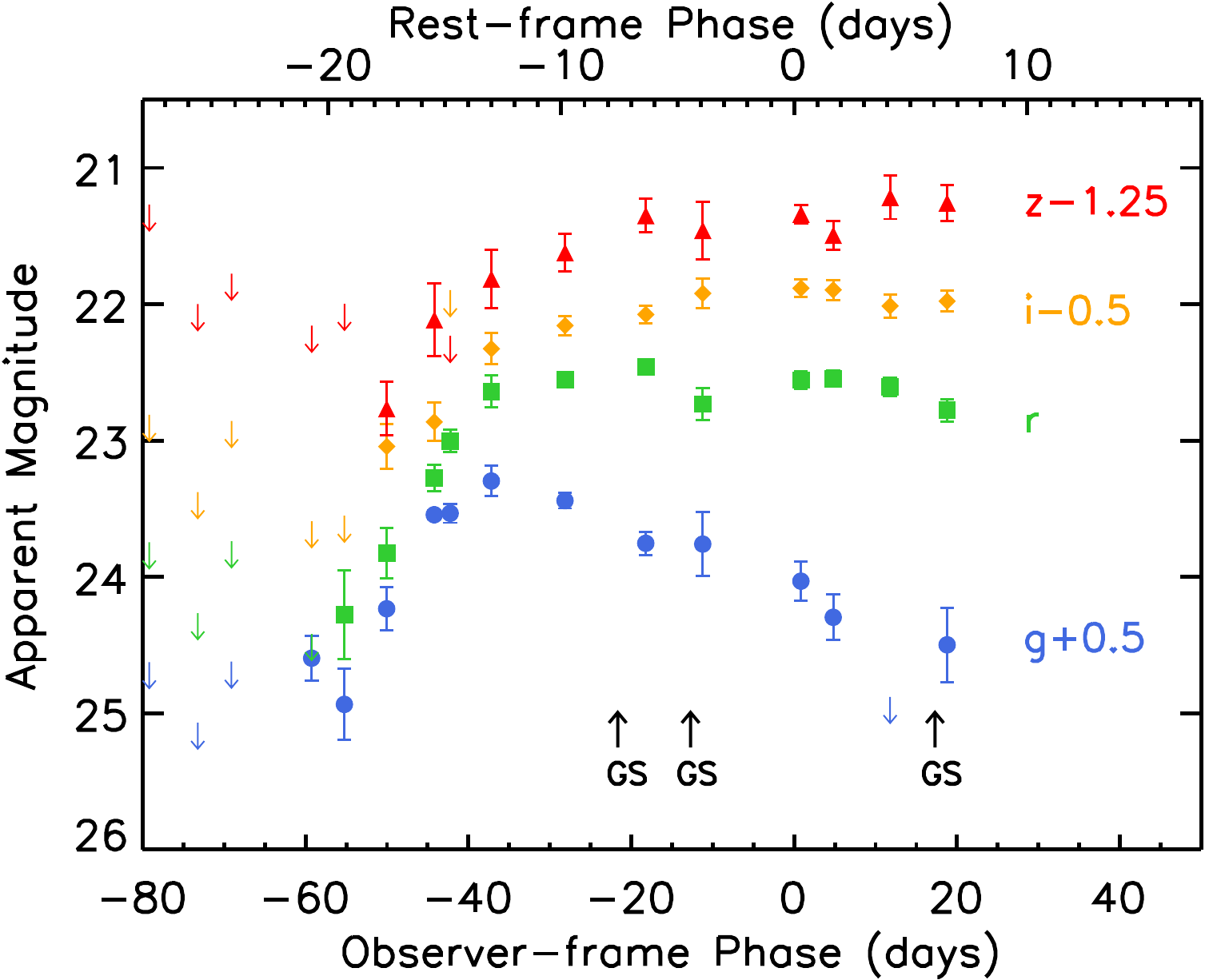}\\
		\includegraphics*[scale=0.41]{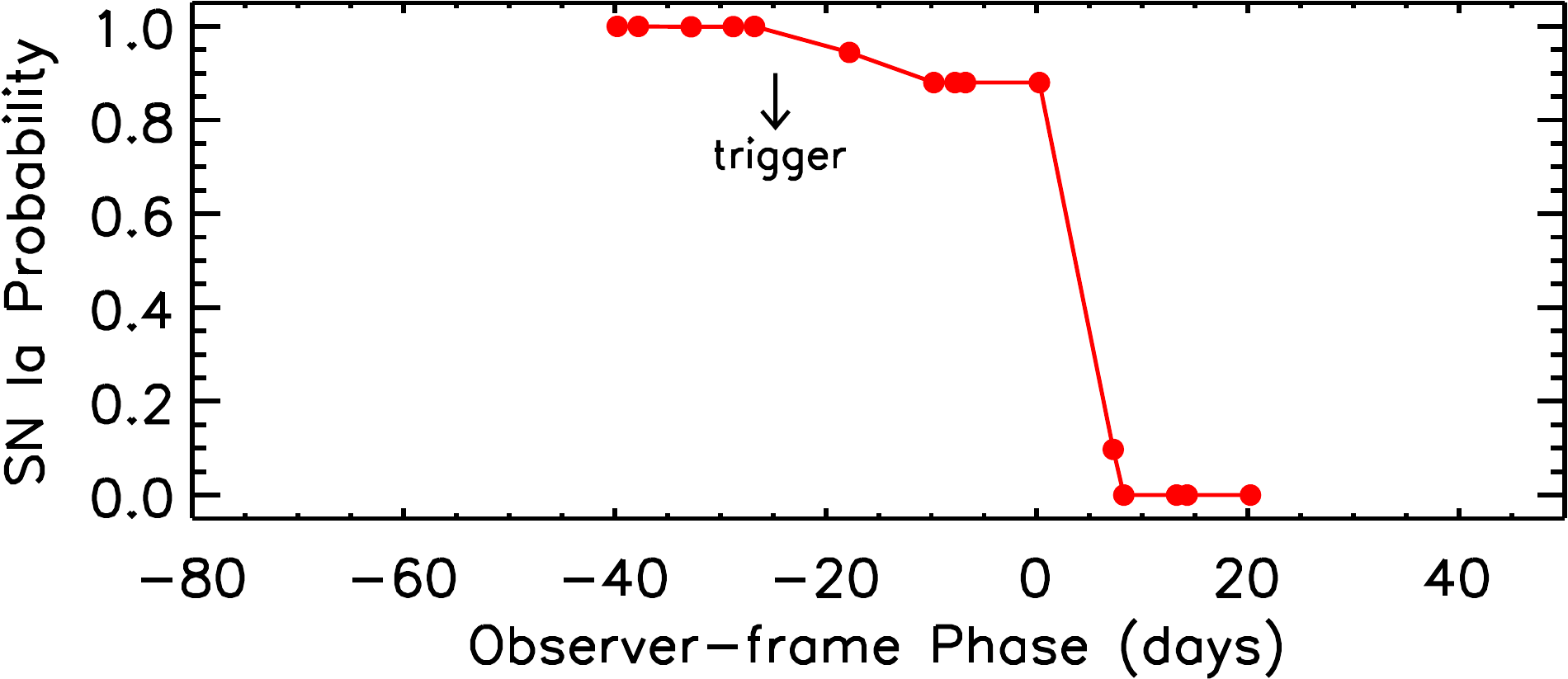}
	\end{tabular}
        \caption{ {\it Top}: DECam $griz$ light curves of DES15E2mlf.
          The phase of observation is relative to the peak of the $i$-band
          light curve on ${\rm MJD} = 57384.26$.
          Upper limits are represented by the downward arrows.  The
          upward arrows in black mark the epochs of Gemini-South
          spectra. {\it Bottom}: {\tt pSNID}
          \citep{2008AJ....135..348S, 2011ApJ...738..162S} probability
          of DES15E2mlf being a SN~Ia as a function of phase. The
          arrow indicates the epoch when we triggered the first
          spectrum.}
            \label{lc}
\end{figure}

DES15E2mlf was discovered in exposures taken by DES on November 7, 2015 (UT) at $\rm
\alpha=00^{h}41^{m}33^{s}.40$, $\delta=-43\degree27'17''.20$
\citep{2015ATel.8460....1P}. It is located in the `E2' shallow field,
one of the 10 SN fields (2 deep and 8 shallow fields) in DES-SN. 
The SN was detected with the difference imaging pipeline
\citep[DiffImg;][]{2015AJ....150..172K}.  The light curves were
observed in DECam \citep{{2015AJ....150..150F}} $griz$ filters 
over $\sim$2 months.  DES15E2mlf was
followed by DES through the end of the observing season \citep{2016SPIE.9910..99101D}
in February 2016.  The DECam images were reduced using the same photometric
pipeline as described in \citet{2015MNRAS.449.1215P} and
\citet{2016ApJ...818L...8S}.  We subtract a deep template image from
each SN image to remove the host-galaxy light using a
point-spread-function (PSF) matching routine.
We then measure the SN photometry by fitting its PSF from the difference image.
The light curves of DES15E2mlf are shown in Fig.~\ref{lc}.

At discovery, the SN was $\sim$1.2\,mag below peak in the $i$ band.
From discovery, the SN rose in all bands.
In fact, the SN was discovered 49 days (17 rest-frame
days) before peak (see Section~\ref{sec:lcs}).  Despite its proximity
to its host galaxy (offset by 0.1\arcsec\ from the nucleus), its
relatively bright luminosity ($\sim$1\,mag brighter than its host)
made DES15E2mlf a good candidate for spectroscopic observations.

Intriguingly, DES15E2mlf was not initially flagged as a potential
SLSN.  Rather, its early light curves were similar to SN~Ia light
curves at $z \approx 0.3$.  Complicating the proper identification was
the fact that its host galaxy has a photometric redshift of $0.343 \pm 0.103$
estimated from the DECam $griz$ photometry \citep{2016PhRvD..94d2005B}, 
consistent with the implied SN redshift.
However, the photometric redshift is way off from the redshift measured from the
host features in the SN spectrum (see Section~\ref{sec:redshift}).  
We found this incorrect photometric redshift of the host galaxy is likely
due to the lack of rest-frame optical photometry 
(see Section~\ref{sec:host} for more details).

In the bottom panel of Fig.~\ref{lc}, we display the probability that
DES15E2mlf was a SN~Ia as determined by the Photometric SN
IDentification software \citep[{\tt pSNID};][]{2008AJ....135..348S,
  2011ApJ...738..162S} as a function of phase.  It was consistently
considered a likely SN~Ia until after peak brightness when the colors
became inconsistent with that of a SN~Ia.  Although DES15E2mlf was 
considered highly unlikely a SN~Ia at that point, {\tt pSNID} was unable 
to classify it as a SLSN due to the lack of SLSN templates.

Based on its initial classification as a likely SN~Ia with a SN
redshift consistent with its host-galaxy photometric redshift, we
triggered spectroscopic follow-up observations through our SN~Ia
program with Gemini-South (program GS-2015B-Q-7; PI Foley).  This
first spectrum was obtained on December 6, 2015. The observations
consisted of $3\times1200$\,s exposures taken with the Gemini
Multi-Object Spectrograph \citep[GMOS;][]{2004PASP..116..425H} using
the R400 grating and a central wavelength of 750\,nm (observed
wavelength range of $\sim$5200--9900\,\AA).  We selected an angle to
place the center of slit (with 1\arcsec\ slit width) through both the SN and the host-galaxy nucleus.

The data reduction process included standard CCD processing and
spectrum extraction with IRAF\footnote{The Image Reduction and
  Analysis Facility ({\tt IRAF}) is distributed by the National
  Optical Astronomy Observatories, which are operated by the
  Association of Universities for Research in Astronomy, Inc., under
  cooperative agreement with the National Science Foundation.}. We
applied our own IDL routines to flux calibrate the data and remove
telluric lines using the well-exposed continua of the
spectrophotometric standard stars \citep{1988ApJ...324..411W,
  2003PASP..115.1220F}.  Details of the spectroscopic reduction are
described in \citet{2012MNRAS.425.1789S}.

\begin{figure}
	\centering
	\begin{tabular}{c}
		\includegraphics*[scale=0.52]{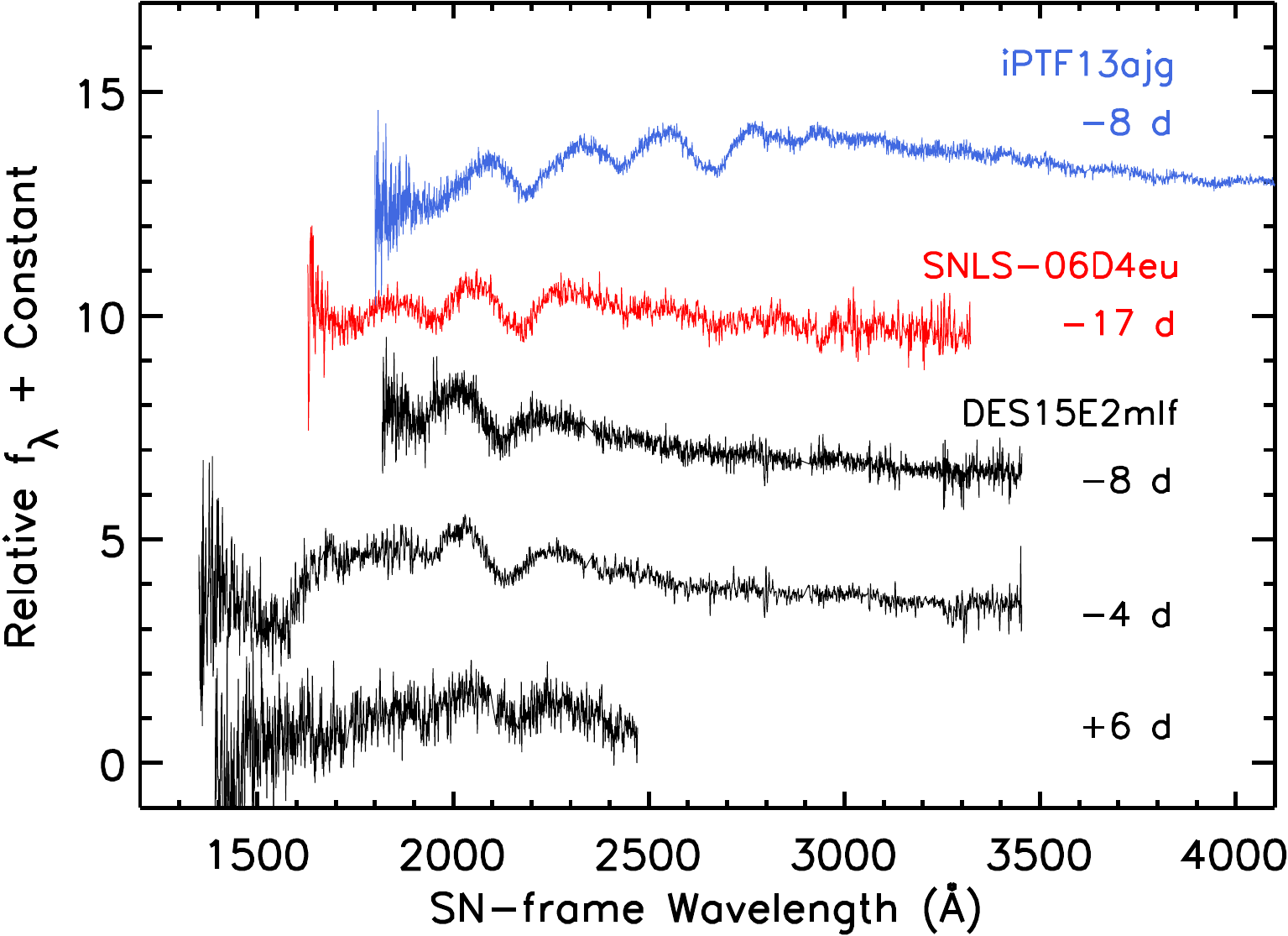}
	\end{tabular}
        \caption{ Gemini-South spectra of DES15E2mlf (black) at $-8$,
          $-4$ and $+6$ days relative to peak $i$-band brightness.  We
          also display the spectra of iPTF13ajg (blue) and SNLS-06D4eu (red), 
          SLSNe-I at $z = 0.740$ and 1.588, respectively. } 
        \label{spec}
\end{figure}

To investigate the potential spectral evolution, two more Gemini/GMOS 
spectra were obtained on December 15, 2015 (program GS-2015B-Q-8; PI Galbany) 
and January 14, 2016 (program GS-2015B-Q-7; PI Foley), respectively.  
The December observations used both
the R400 configuration and an additional configuration with the B600
grating (observed wavelength range of $\sim$3800--7000\,\AA) for a
total wavelength range of $\sim$3800--9900\,\AA.  The January spectrum
only utilized the B600 grating.  Two additional spectra were obtained
with the AAOmega Spectrograph \citep{2004SPIE.5492..389S} and the 2dF
fibre positioner at the Anglo-Australian Telescope (AAT) on December
12 and 14. However, other than noting that the continuum of 
DES15E2mlf was conspicuously blue, the signal-to-noise ratio (S/N) of the AAT spectra
are too low (${\rm S/N} \lesssim 1$) to confidently identify either SN
or host galaxy features. Thus, in this work we only present an
analysis of the Gemini spectra. The whole spectral sequence and log of
spectroscopic observations can be found in Fig.~\ref{spec} and
Table~\ref{spec-log}, respectively.

\begin{table}
\centering
\caption{Log of spectroscopic observations of DES15E2mlf.}
\begin{tabular}{lccc}
\hline\hline
Date & Instrument & Grating & Exp. time\\
(MJD)&            &         &    (s)\\
\hline
57362.59 & Gemini-S/GMOS$^a$ 	& R400 & 3600\\
57371.53 & Gemini-S/GMOS$^b$ 	& B600/R400 & 3600/1800\\
57401.55 & Gemini-S/GMOS$^a$ 	& B600 & 4800\\
\hline
\end{tabular}
\label{spec-log}
\\$^a${The spectra were observed under GS-2015B-Q-7 (PI Foley).}
\\$^b${The spectra were observed under GS-2015B-Q-8 (PI Galbany).}
\end{table}

\section{Data and Analysis}
\label{sec:analysis}

\subsection{Redshift}
\label{sec:redshift}
\begin{figure*}
	\centering
	\begin{tabular}{c}
		\includegraphics*[scale=0.9]{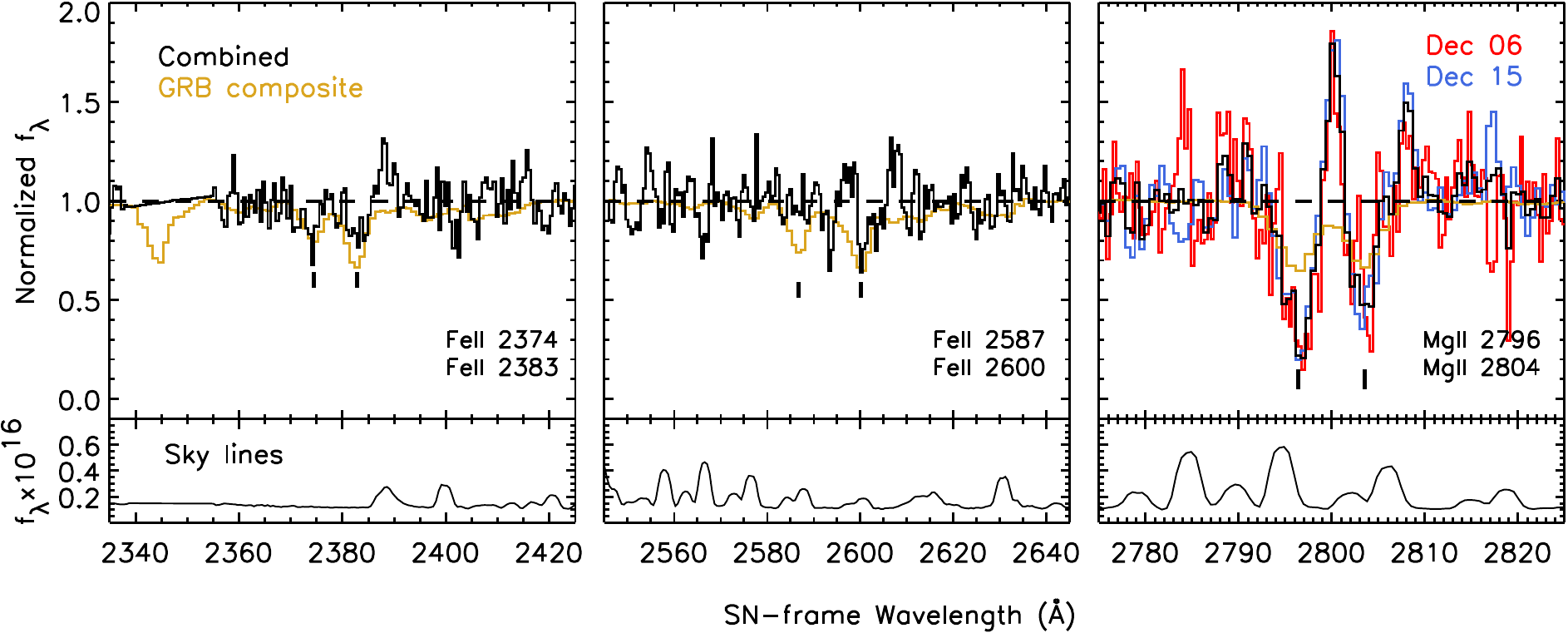}
	\end{tabular}
        \caption{
          Upper panel: Continuum-normalized spectra of DES15E2mlf near the \Feii\ 
          (left and middle panels) and \mgii\ (right panel) doublets.  Each line
          is marked.  The December 6 and December 15 spectra are shown
          as blue and red curves, respectively.  The combined spectrum
          is shown in black.  The GRB composite spectrum from \citet{2011ApJ...727...73C}
          is shown in gold. Lower panel: The spectrum of extracted sky emission lines. }
        \label{absorb}
\end{figure*}

In Fig.~\ref{spec}, we present the Gemini spectra of DES15E2mlf at
phases of $-8$, $-4$ and $+6$ days (in the rest-frame) relative to
$i$-band maximum brightness.
There are clear narrow absorption features in the spectra of
DES15E2mlf.  We identify the most prominent lines as the \mgii\
$\lambda\lambda 2796$,2804 doublet at $z = 1.8607 \pm 0.0003$ (see Fig.~\ref{absorb}).  
We find no significant evolution of the absorption lines for the phases
covered by our spectra. 
Following \citet{2012ApJ...755L..29B}, we
compare the continuum-normalized spectra of DES15E2mlf to the
\citet{2011ApJ...727...73C} GRB composite spectrum
in Fig.~\ref{absorb}. Also marginally detected ($\sim2\sigma$ significance) in the spectra are
\Feii\,$\lambda\lambda2374,2383$ and \Feii\,$\lambda2600$
lines, although they are much weaker than the \mgii\
doublet. There are no other obvious absorption features at other
redshifts. At this redshift, DES15E2mlf is also consistent with that of hydrogen-poor SLSNe-I
(see Section~\ref{sec:spec}).

\subsection{Classification}
\label{sec:class}
Examining the spectra, we find that DES15E2mlf is most similar to hydrogen-poor SLSNe-I.
However, our spectra do not cover the wavelengths of any hydrogen
lines, and therefore, we cannot definitively rule out the possibility
of DES15E2mlf being a SLSN-II.  Nonetheless, based on the spectra
alone, the best classification is that of a SLSN-I.

Further supporting the SLSN classification is its peak luminosity.
DES15E2mlf had a peak apparent AB magnitude of $i = 22.4\pm0.1$\,mag
($\lambda_{\rm rest}\approx2745$\,\AA).  Correcting for minimal Milky
Way extinction \citep[$A_{i} = 0.02$\,mag;][]{2011ApJ...737..103S} and
cosmic expansion, DES15E2mlf peaks at $M_{\rm AB} = -22.3\pm0.1$\,mag.  At this
luminosity, DES15E2mlf is well above the (arbitrary) SLSN threshold
\citep[$M < -21$\,mag;][]{2012Sci...337..927G}.

\citet{2012Natur.491..228C} reported two SLSNe at $z=2.05$ and
3.90, respectively. However, the lack of solid spectroscopic observations 
made the classification uncertain and questionable. At $z = 1.861$, DES15E2mlf is the most 
distant spectroscopically confirmed SLSN which exploded only 3.5\,Gyr
after the Big Bang.

\subsection{Spectra}
\label{sec:spec}
The multiple spectra of DES15E2mlf show two strong absorption features
at rest-frame wavelengths of $\sim$1950 and 2150\,\AA, respectively.
No strong features are found between 2300 and 3500~\AA.  The
spectra are similar to those of other SLSNe, and in particular to that
of SNLS-06D4eu (see Fig.~\ref{spec}), however, DES15E2mlf has the absorption features
shifted further to the blue, with even the latest spectrum of
DES15E2mlf being further to the blue by $\sim1800$\,km\,$\rm s^{-1}$ than the $-17$-day spectrum of
SNLS-06D4eu. However, it is still smaller than the typical expansion velocity a SLSN may have
\citep[$\rm\sim10\,000-20\,000\,km\,s^{-1}$; e.g.,][]{2013ApJ...770..128I,2016MNRAS.458.3455M}.
Here the velocity is calculated by firstly deredshifting and smoothing the spectrum using 
an inverse-variance-weighted Gaussian filter. We then use the wavelength of the maximum absorption in the smoothed spectrum to calculate the velocity \citep[see][for more details]{2006AJ....131.1648B}.
\citet{2016MNRAS.458.3455M} identified the absorption
features as \Feiii\ and \ciii, respectively. In Fig.~\ref{spec} we
also compare the spectrum to that of iPTF13ajg, a SLSN-I at
$z=0.74$. We found both high-redshift DES15E2mlf and SNLS-06D4eu lack some
spectral features (or they could be relatively weak) seen in iPTF13ajg, particularly the $2700$~\AA\
line (mostly resulting from \mgii), which was commonly found in many of low-redshift
SLSNe. 

\begin{figure}
	\centering
	\begin{tabular}{c}
		\includegraphics*[scale=0.52]{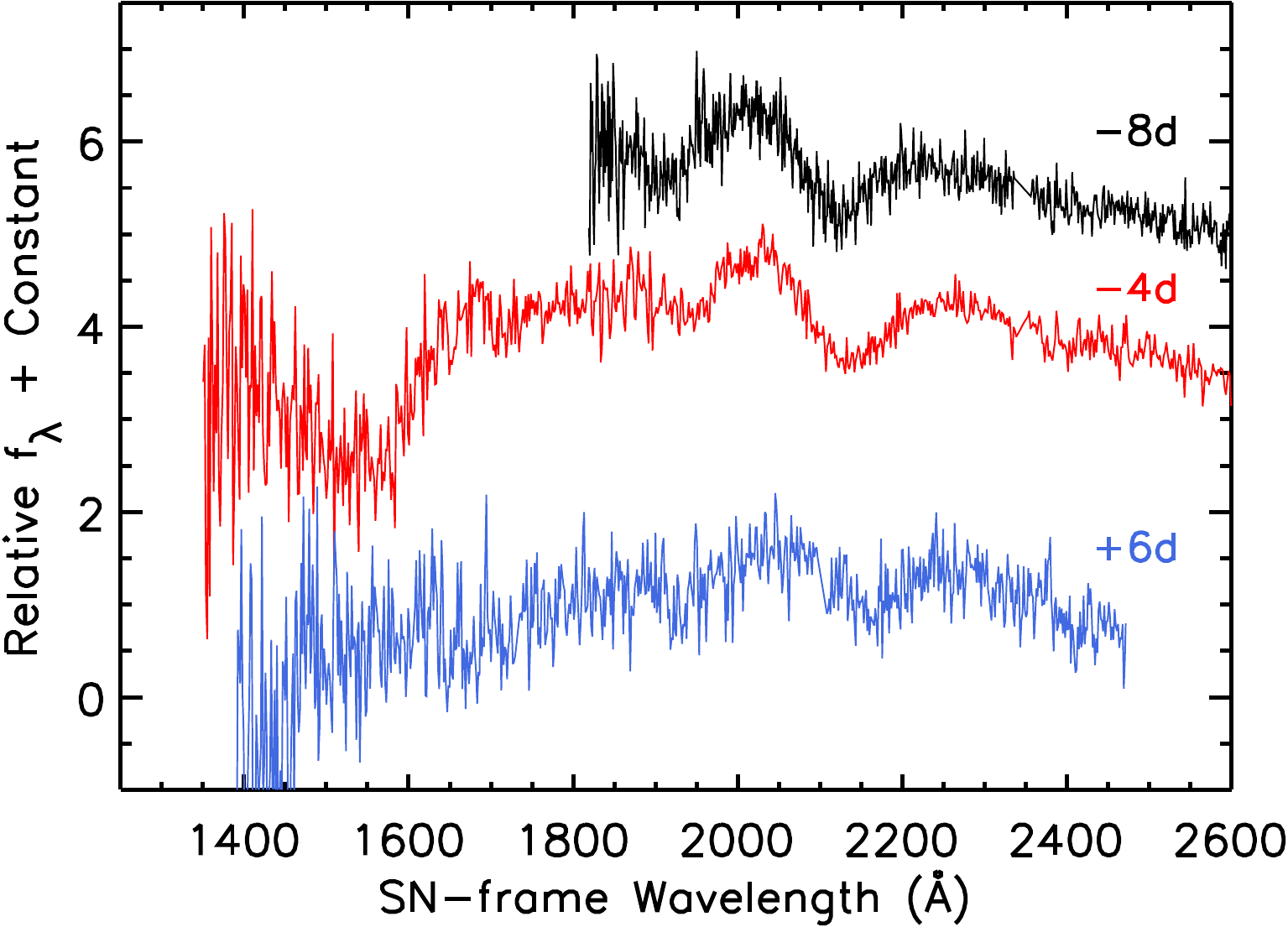}\\
		\includegraphics*[scale=0.67]{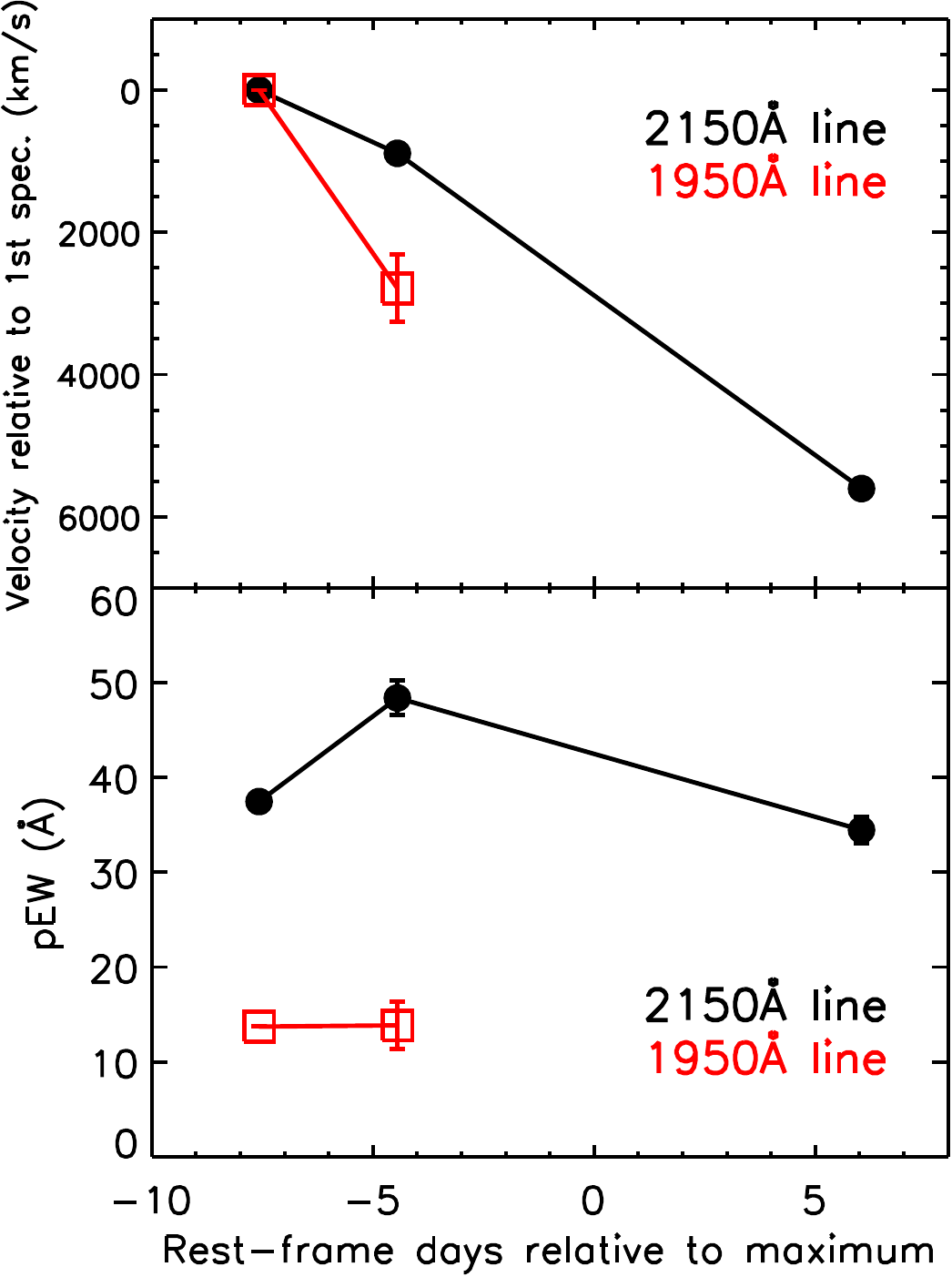}
	\end{tabular}
        \caption{{\it Top}: Spectra of
          DES15E2mlf focusing on the two strong absorption features at
          $\sim$1950\,\AA\ and 2150\,\AA.  {\it Middle}: Velocity of the
          2150\,\AA\ (black circles) and 1950\,\AA\ (red open squares)
          absorption lines relative to the velocity in the first epoch
          as a function of phase.  {\it Bottom}: Pseudo-equivalent
          width (pEW) for the two features as a function of phase.}
        \label{spec-uv}
\end{figure}

Given the high redshift of DES15E2mlf, our multiple optical
spectra probe the far-UV spectral evolution of a SLSN. 
In Fig.~\ref{spec-uv}, we display the spectra of
DES15E2mlf from $-8$ to $+6$~days relative to peak brightness,
focusing on the absorption features identified above. Here, we
investigate the velocity evolution of DES15E2mlf, but we describe any
velocity changes relative to the first spectrum so as to not assume a
particular rest-frame wavelength. 

Over the time covered by our spectra, the
$\sim$2150\,\AA\ feature (identified as \ciii)
was redshifted by 5600\,km\,s$^{-1}$.  The $\sim$1950\,\AA\ feature
(identified as \Feiii) is not strongly detected in our third spectrum.
However, between the first two epochs ($-8$ and $-4$~days), the
feature was redshifted by 2800\,km\,s$^{-1}$.   
We present this evolution in Fig.~\ref{spec-uv}.
A similar trend was found in other SLSNe \citep[e.g., iPTF13ajg;][]{2014ApJ...797...24V}.
Assuming the 2150\,\AA\ feature is predominantly caused by
the \ciii\ line (with a rest wavelength of 2297\,\AA), we calculate
a expansion velocity of $\sim22\,600$, 21\,800 and 17\,400\,km\,s$^{-1}$ at
phases of $-8$, $-4$ and $+6$ days, respectively. This velocity evolution is consistent
with those found from other SLSNe.

While the velocity of the strong absorption features changed
significantly during the time period covered by our spectral
observations, no significant change is found in the equivalent widths
of those features (Fig.~\ref{spec-uv}). However, we are aware that there is possibly
some host-galaxy contamination on our spectra which could affect
the velocity and equivalent width measurement. This effect
should be small given that our spectra were observed when the SN
light was still dominant (more than 1\,mag brighter than the host galaxy).
We test this by determining the best-fit SED model of the host galaxy
given by {\tt Z-PEG} (see Section~\ref{sec:host}) and subtracting that 
from the SN spectrum. We found the host-galaxy contamination has negligible effect
on our results.

\subsection{Light curves}
\label{sec:lcs}
In Fig.~\ref{lc}, we present the DECam $griz$ light curves of
DES15E2mlf (which have central wavelengths corresponding to $\lambda_{\rm
  rest} = 1700$, 2259, 2745 and 3215\,\AA\ in the rest frame,
respectively). We fit a polynomial to each light curve to determine
the peak luminosity and time of maximum brightness.  We find that the
SN had a peak $i$-band magnitude of $22.38\pm0.07$\,mag on ${\rm MJD} = 57384.26\pm0.08$.
Here the uncertainties of peak magnitude and MJD are computed by simply
propagating the 1-$\sigma$ error of the $i$-band magnitude in the polynomial fit,
and assuming the epochs of those measurements all have zero errors.
The light curve rose for 62 observer-frame days to peak (in $i$-band), 
corresponding to a rest-frame rise time of
21.5~days, comparable to that of SNLS-06D4eu ($\sim30$~days).  
The time of maximum is later for redder bands.  That is, the $grz$
light curves peak $-35$, $-23$, and $+3$ observer-frame days ($-12$,
$-8$, and $+1$ rest-frame days) relative to that of $i$-band,
respectively.

While the $riz$ light curves have a similar appearance, the $g$-band
light curve evolves differently.  Specifically, the $g$ band (rest
frame 1700~\AA) reaches peak brightness much earlier than the other
filters and then declines much faster than the other bands, decreasing
$\sim$1\,mag in 14 rest-frame days.  The $g$-band light curve of
DES15E2mlf probes bluer wavelengths than most previous SLSN-I light
curves. Its rapid evolution compared to other bands may
suggest unique information (e.g., SN cooling) at these wavelengths.
However, the lack of early-time spectra (in $g$-band) prevents us
from making an explicit interpretation on our results.

\begin{figure}
	\centering
	\begin{tabular}{c}
		\includegraphics*[scale=0.5]{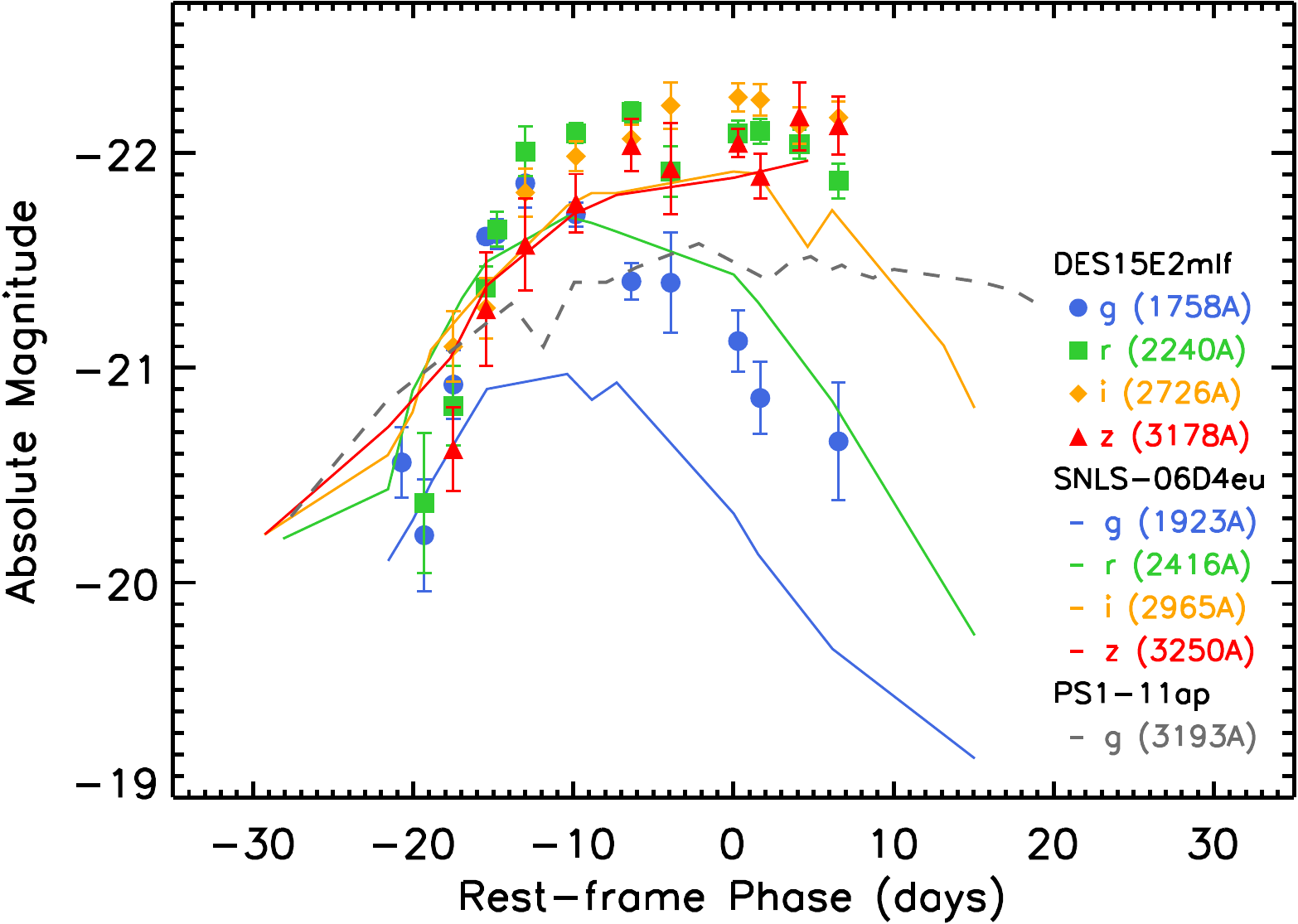}
	\end{tabular}
        \caption{Absolute-magnitude light curves of
          DES15E2mlf (solid symbols), the $z = 1.588$ SLSN-I
          SNLS-06D4eu \citep[solid lines;][]{2013ApJ...779...98H}
          and the $z = 0.524$ SLSN-I
          PS1-11ap \citep[dashed lines;][]{2014MNRAS.437..656M}.
          The $griz$ bands are labeled with their rest-frame
          effective wavelength for each SN. }
        \label{lc-comp}
\end{figure}

In Fig.~\ref{lc-comp}, we compare the absolute magnitude light curves
of DES15E2mlf with those of SNLS-06D4eu, a SLSN at $z = 1.588$
\citep{2013ApJ...779...98H}.  The absolute magnitudes for both SNe
are determined by only correcting for the Milky Way extinction and cosmological expansion
\citep[e.g., Eq.~1 in][]{2014MNRAS.437..656M}. No proper $K$-correction
is applied here. Since DES15E2mlf and SNLS-06D4eu have
somewhat similar redshifts, the rest-frame effective wavelength for the
$griz$ filters are similar, making direct comparisons reasonable and simpler.
Also compared is the $g$-band light curve of PS1-11ap, a well studied SLSN
at lower redshift \citep[$z=0.524$;][]{2014MNRAS.437..656M}. The $g$ filter of PS1-11ap 
has a similar rest-frame wavelength (3193~\AA) as that for the $z$ filter of DES15E2mlf,
providing a good comparison to our near-$UV$ data.

DES15E2mlf has higher peak luminosities than those for SNLS-06D4eu at all bands, with
the largest difference of 0.9~mag in $g$.  The two SNe evolve similarly in
$i$ and $z$ (corresponding to roughly 2850 and 3210~\AA\ in the rest
frame) for the range where both SNe have data.  However, DES15E2mlf
peaks later and has a slower decline in $g$ (\about 1840~\AA\ in the
rest frame) and $r$ (\about 2330~\AA\ in the rest frame).  While the
exact timing and decline rate are different for the two SNe in $g$,
both have the same basic behaviour of peaking earlier and declining
faster in $g$ than in the other bands.  The combination of a slower
rise, slower decline, and generally higher peak luminosity in each
band means that the integrated luminosity was larger for DES15E2mlf
than SNLS-06D4eu (at least over the time when both SNe have data).
The lower-redshift PS1-11ap shows similar evolution in $g$ 
(matches the $z$-band rest-frame wavelength of DES15E2mlf and SNLS-06D4eu),
but has the lowest peak luminosity among all the SNe ($\sim0.5$\,mag lower than that of DES15E2mlf).

\subsection{Host galaxy}
\label{sec:host}

\begin{figure}
	\centering
	\begin{tabular}{c}
		\includegraphics*[scale=0.4]{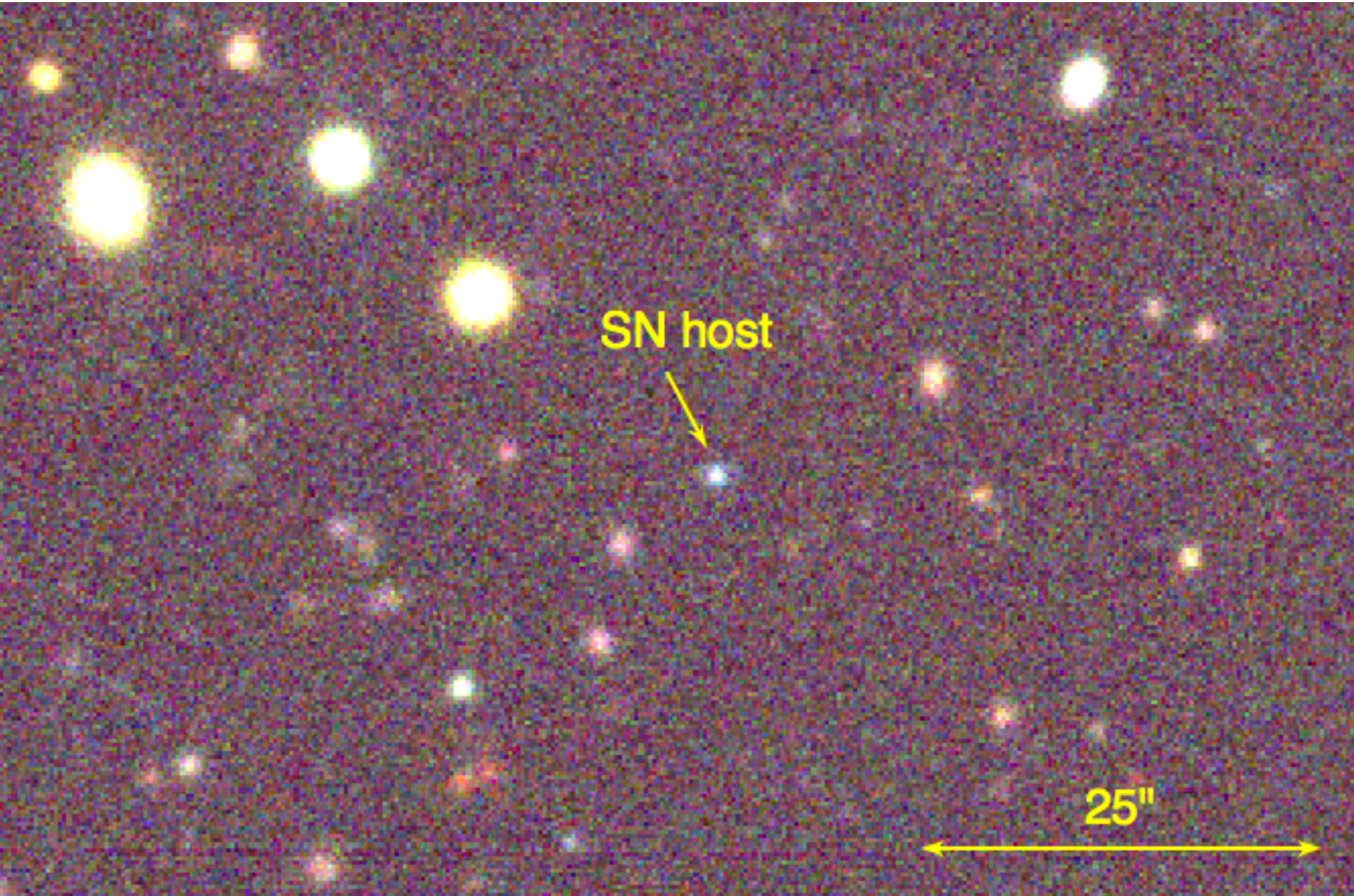}
	\end{tabular}
     \caption{False-color (RGB channels corresponding to DECam $irg$ filters) 
     template image of the field surrounding DES15E2mlf, with the yellow arrow marking the 
     SN host galaxy. North is up and east is left.  The host is noticeable blue compared 
     to other objects in the field.
     }
     \label{stamp}
\end{figure}

Previous studies have shown that hydrogen-poor SLSNe tend to reside in
low-mass galaxies with strong star formation, with the current sample
of 31 SLSN-I host galaxies having an average stellar mass (\mstellar)
of $\log\rm(M/M_{\sun}) \approx 8.30$ and a specific star-formation
rate (sSFR) of $\approx\rm 2\,Gyr^{-1}$ \citep{2014ApJ...787..138L}.
Similar results were found in \citet{2016arXiv160408207P} with
a lower-$z$ SLSN-I sample. 
Here we examine the properties of the host galaxy of DES15E2mlf.

Using DES images of the host galaxy of DES15E2mlf, we measure $griz$
AB magnitudes of $23.46\pm0.04$, $23.30\pm0.05$, $23.47\pm0.08$, and
$23.52\pm0.11$\,mag, respectively. Fig.~\ref{stamp} shows a DES composite image
of the host galaxy and its surrounding field. We also obtain NIR photometry of 
the host galaxy from VISTA Deep Extragalactic Observations \citep[VIDEO;][]{2013MNRAS.428.1281J}. 
We measure $ZYJHK_{s}$ AB magnitudes of 23.50$\pm$0.05, 23.56$\pm$0.06\,mag, 23.31$\pm$0.11\,mag,
23.56$\pm$0.10\,mag, 23.62$\pm$0.20\,mag, respectively. The addition of NIR 
photometry is critical for setting more robust constraints on the host-galaxy 
parameters compared to those obtained from the optical data alone.

Using the photometric redshift
code {\tt Z-PEG} \citep{2002A&A...386..446L} and assuming a
\citet{1955ApJ...121..161S} initial-mass function (IMF), we measure a
stellar mass of $\log\rm (M/M_{\sun}) = 9.54^{+0.31}_{-0.50}$, which is more
massive than the typical SLSN-I host galaxy. The photometry also
indicates a specific star-formation rate (sSFR) of $\rm2.46^{+0.00}_{-1.01}\,Gyr^{-1}$.
We also notice that the photometric redshift of the host galaxy estimated by {\tt Z-PEG} 
changed from 0.2 to 1.9 after including the VIDEO data, 
which is now consistent with the spectroscopic redshift of the SN. 
Therefore this deviation is likely 
due to the lack of rest-frame optical photometry
(note that the DECam photometry of DES15E2mlf are all in rest-frame UV) 
in estimating the initial photometric redshift of host galaxy.

\begin{figure}
	\centering
	\begin{tabular}{c}
		\includegraphics*[scale=0.48]{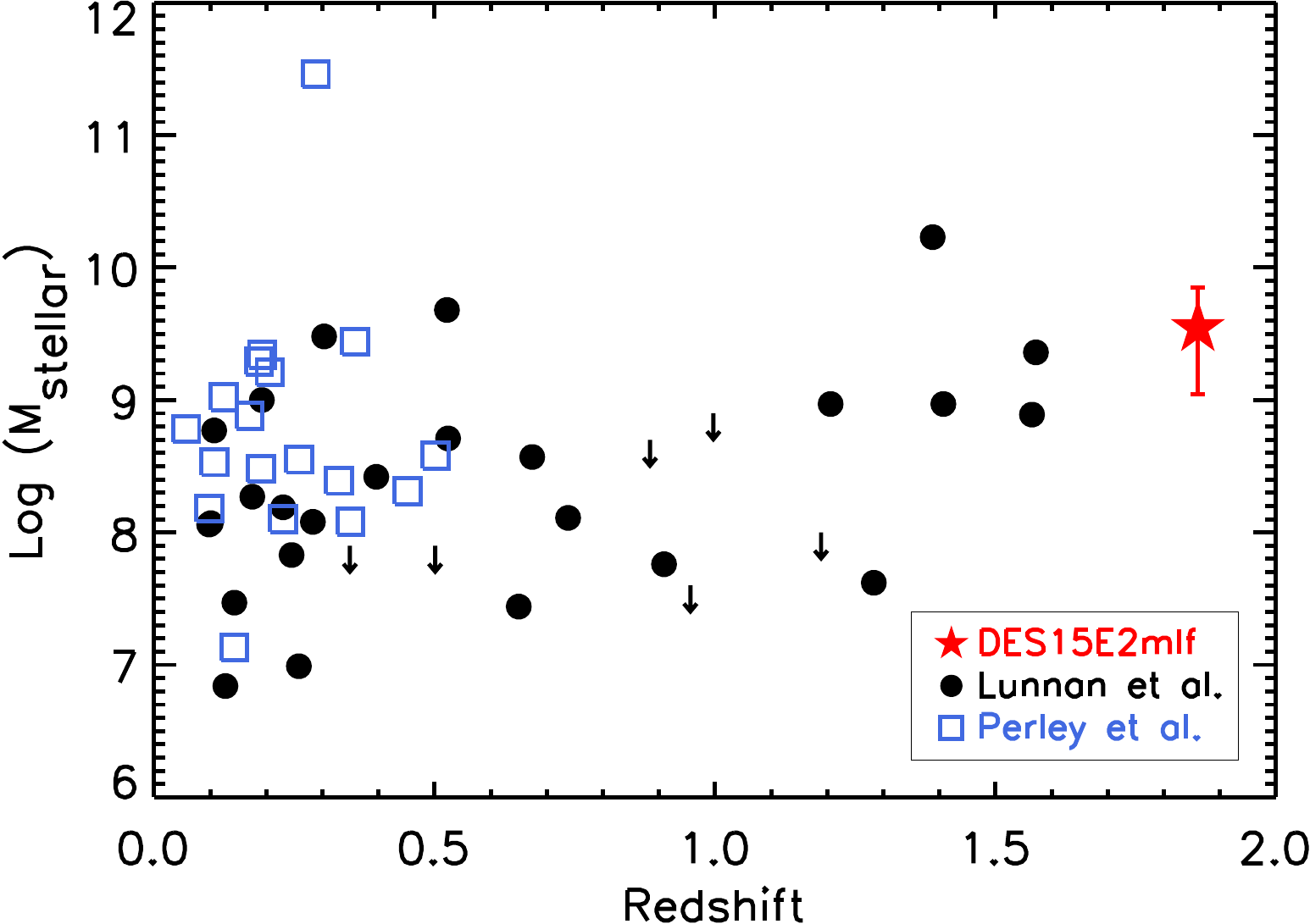}\\
		\includegraphics*[scale=0.48]{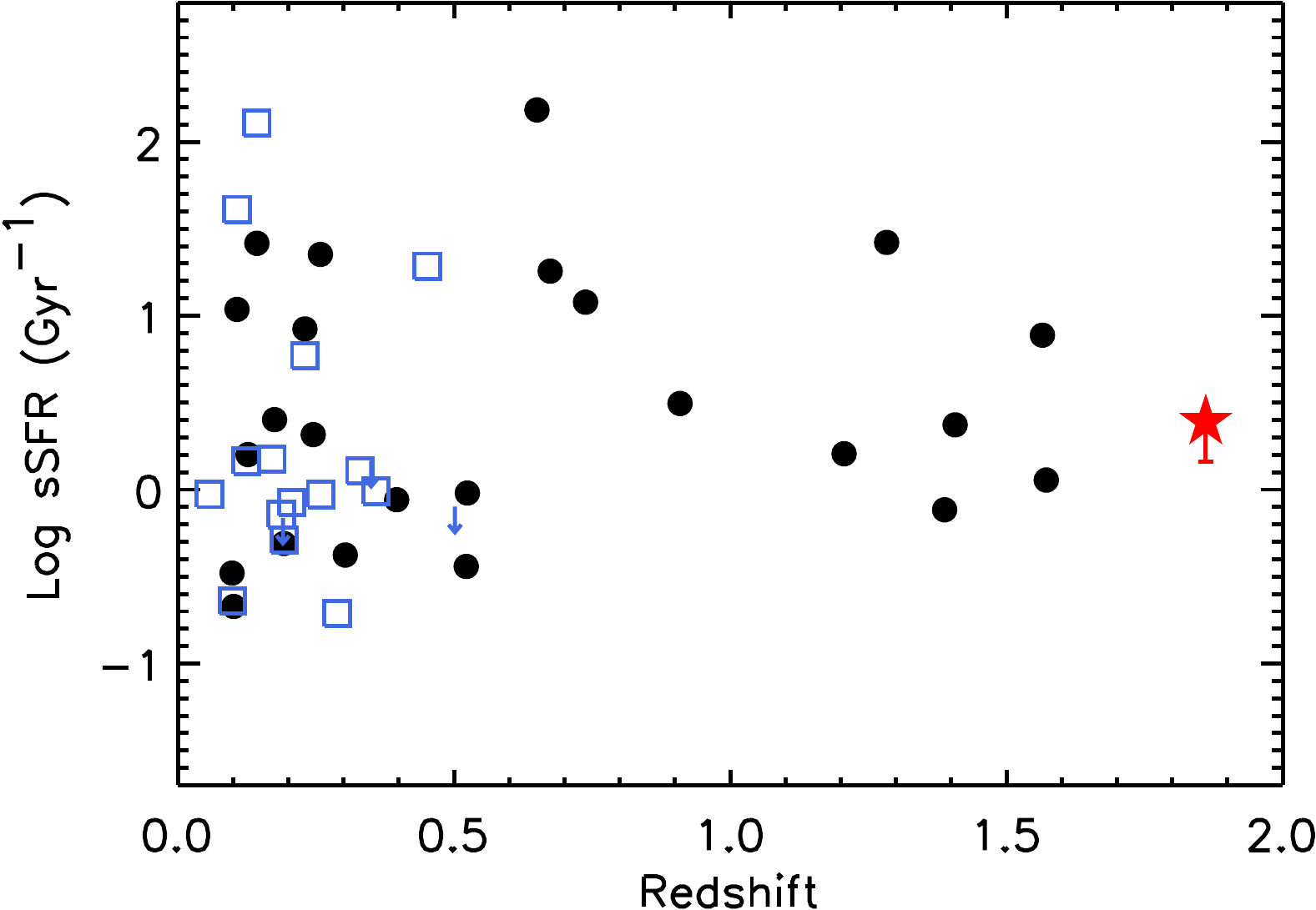}\\
	\end{tabular}
        \caption{ SLSN-I host-galaxy
          \mstellar\ as a function of redshift (top panel) and sSFR as a
          function of redshift (bottom panel).  Black points and blue squares
          represent the \citet{2014ApJ...787..138L} and \citet{2016arXiv160408207P} sample,
          respectively, while the red star represents the host galaxy of DES15E2mlf.  
          }
        \label{hostmass}
\end{figure}

We present these values compared to those of the
\citet{2014ApJ...787..138L} and \citet{2016arXiv160408207P} sample in 
Fig.~\ref{hostmass}. Similar stellar population synthesis techniques were
used to obtain the host parameters in \citet{2014ApJ...787..138L} and \citet{2016arXiv160408207P}.
To be consistent, we convert the host stellar mass and SFR to Salpeter IMF if necessary.

DES15E2mlf has one of the most massive host galaxies
discovered for a SLSN-I.  This continues the trend of increasing stellar
mass (on average) with redshift for the \citet{2014ApJ...787..138L}
sample.  Compared to the same sample, the host galaxy of DES15E2mlf
has a higher \mstellar\ than 94\% (29 out of 31) of all SLSN host galaxies.  
However, we cannot rule out the possibility that the host is a satellite galaxy with lower \mstellar.

\section{Discussion}
\label{sec:discussion}
\subsection{Precursor}
\label{sec:precursor}

\begin{figure*}
	\centering
	\begin{tabular}{c}
		\includegraphics*[scale=0.52]{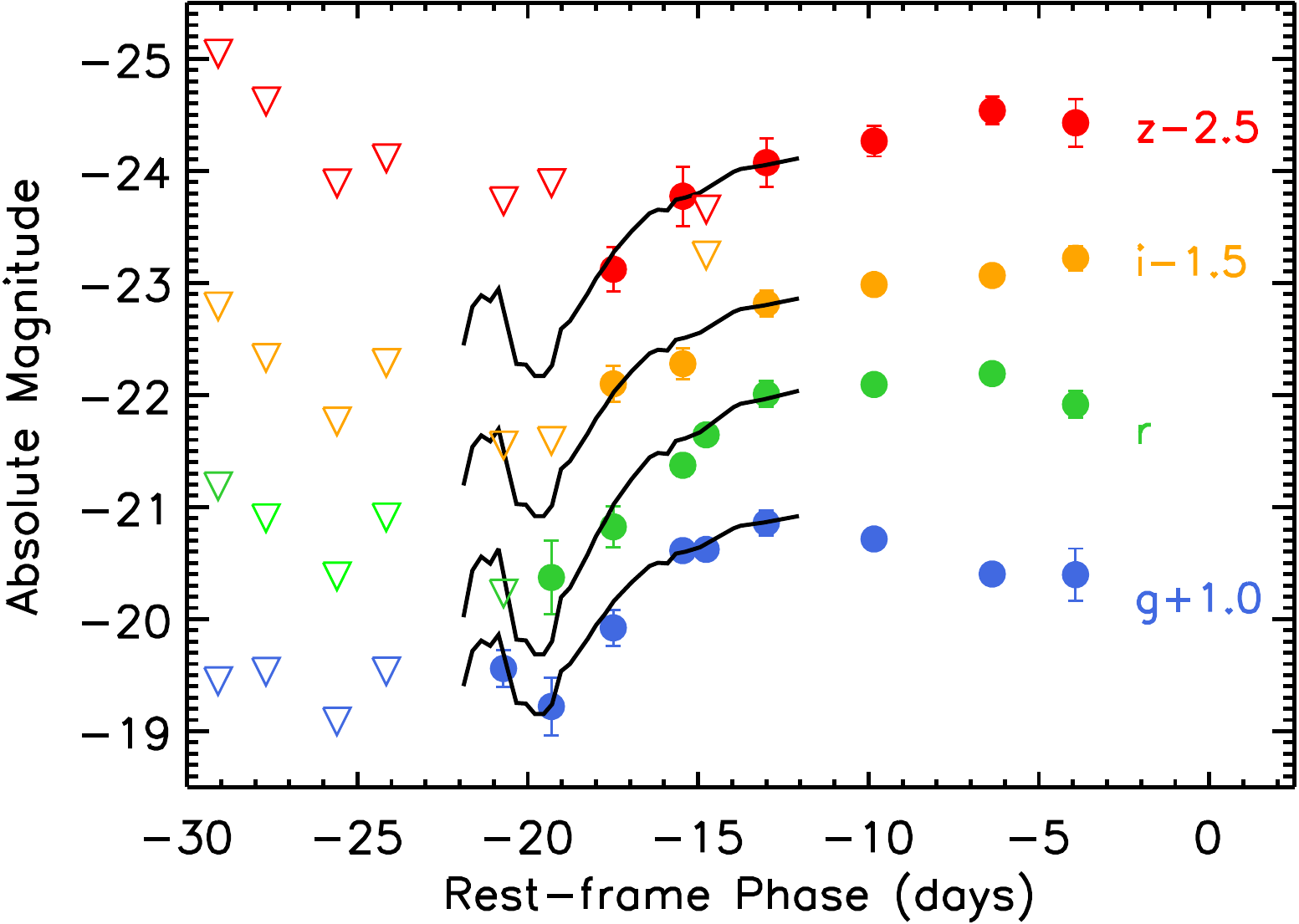}
		\includegraphics*[scale=0.52]{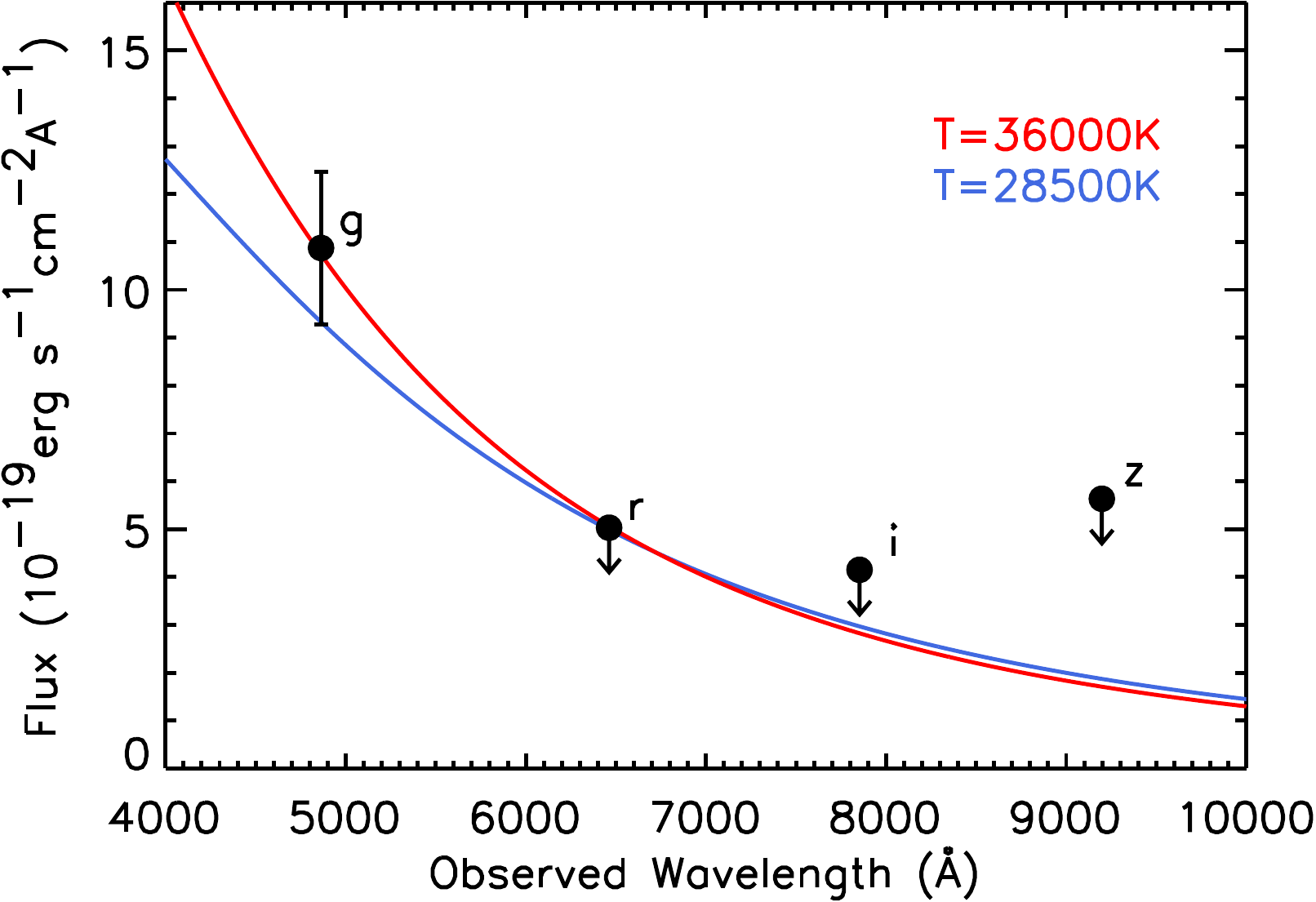}
	\end{tabular}
        \caption{{\it Left}: A closer look at DES15E2mlf pre-maximum light curves. 
        The upper limits are shown in downward triangles.
        We overplot the scaled LSQ14bdq rest-frame $g$-band light curve \citep{2015ApJ...807L..18N} in black.
        {\it Right}: The DES15E2mlf $g$-band flux and upper limits for $riz$ bands at $-21$~days.
        The blackbody curves with $T=$ 28,500\,K (blue) and 36,000\,K (red) are overplotted.}
        \label{precursor}
\end{figure*}

Examining the light curves in detail, we notice that the first detection 
in $g$, at $-21$~rest-frame days relative to peak brightness, is $\sim$1\,mag
brighter than that expected from an extrapolation of the later data.
We do not detect any excess flux in other bands at this epoch.
A difference of $0.34\pm0.31$\,mag is calculated between the first and
second detections in $g$. While this deviation is insignificant ($\sim1\sigma$) 
given the photometric uncertainties, it is also potentially consistent with luminous precursor
events associated with other SLSNe \citep[e.g.,][]{2012A&A...541A.129L,
2013ApJ...779...98H, 2015ApJ...807L..18N, 2016ApJ...818L...8S}.
These peaks are thought to be related to some shock-heating scenarios 
\citep[e.g.,][]{2015ApJ...808L..51P,2016ApJ...821...36K}, which could
provide us critical information of SN progenitor and 
explosion physics. 

Following \citet{2016MNRAS.457L..79N}, we use the rest-frame $g$-band
light curve of LSQ14bdq (which has a well-sampled precursor) as a
template to test the existence of a potential precursor for DES15E2mlf 
(see Fig.~\ref{precursor}). The template is arbitrarily scaled by
visual inspection to match the DES15E2mlf $g$-band light curve. For
the $riz$ bands, we apply the same temporal scaling factor to the
template as in the $g$ band, but arbitrarily scale the flux of the
template to match the $riz$ light curves.

To match the template to the rising portion of the light curve, 
the first $g$-band detection must occur
during the declining phase of the precursor.  Therefore, because of
the lack of good earlier detections, the peak precursor luminosity is
limited to $\rm M_{1700\AA} < -20.56$\,mag (assuming that the first
detection is of a precursor). For the $riz$ light curves, the templates
are either consistent or below the detection limits.

For the epoch of the first detection, we measure color limits of
$g-r<-0.22$, $g-i<-0.01$, and $g-z<0.67$\,mag.  The $g-r$ color is
particularly constraining for the precursor temperature.  Assuming a
blackbody, we find that the precursor has $T_{bb}\sim$ 36,000\,K,
with a lower limit of $\sim$\,28,500\,K (see also Fig~\ref{precursor}). 
This is comparable to temperatures
measured for the precursors for other SLSNe \citep[e.g., $T_{bb} \simeq
$ 25,000\,$\pm$\,5000\,K;][]{2016MNRAS.457L..79N}.

While our data do not strongly argue for a precursor, they are
consistent with such an event.

\subsection{Interstellar Absorption}
\label{sec:interstellar-absorp}
Quasars, GRBs and SLSNe are powerful tools in probing the interstellar
medium (ISM) of distant galaxies because of their extremely high
luminosities. \citet{2006ApJ...636..610R} studied a sample of quasars that have
intervening \mgii\ absorption systems. They found the fraction of
damped Lyman-$\alpha$ absorbers \citep[DLAs;][]{1986ApJS...61..249W}
increased with \mgii\ equivalent widths (EW). DLAs are characterized by
neutral hydrogen gas clouds with a high column density ($\rm
N_{H\,\textsc{i}} > 2 \times 10^{20}\,cm^{-2}$).
\citet{2006ApJ...636..610R} found that \about 36\% of all systems with
$\rm{EW}(\mgii\,\lambda2796)>0.5$\,\AA\ and $\rm{EW}(\Feii\,\lambda2600)>0.5$\,\AA\
are DLAs. The rate increases to \about 42\% with
$\rm{EW}(\mgii\,\lambda2796)/\rm{EW}(\Feii\,\lambda2600) < 2$ and
$\rm{EW}(\mgi\,\lambda2852)>0.1$\,\AA, and that
essentially no DLAs have
$\rm{EW}(\mgii\,\lambda2796)/\rm{EW}(\Feii\,\lambda2600) > 2$.  
Therefore, measurements of these absorption features can predict the presence of
a DLA.

Examining the narrow absorption lines detected in the spectrum, 
we measure the rest-frame equivalent widths to be
$\rm{EW}(\mgii\,\lambda2796) = 2.75\pm0.04$\,\AA\ and
$\rm{EW}(\mgii\,\lambda2803) = 1.83\pm0.05$\,\AA\ for DES15E2mlf.
However, without the robust detection of $\Feii\,\lambda2600$ and $\mgi\,\lambda2852$
lines, we are not able to constrain an intervening DLA system.

Following \citet{2012ApJ...755L..29B}, we estimate a lower limit on the
column density of $\log N(\mgii) \gtrsim 13.9$.  The
$\mgii$ EW is higher than that for iPTF13ajg \citep[0.5\,\AA;][]{2014ApJ...797...24V},
PS1-11bam \citep[1.3\,\AA;][]{2012ApJ...755L..29B} and GRB composite
spectrum \citep[1.5\,\AA;][]{2011ApJ...727...73C}, and even \about85\%
of the quasar intervening systems \citep[e.g.,][]{2011AJ....141..137Q}.  
These measurements provide possibilities of using distant SLSNe as probes of
ISM from the present all the way back to the early universe.

\subsection{Host galaxy and selection bias}
\label{sec:host2}

We examined the host galaxy of DES15E2mlf, finding it to be higher
mass than typical SLSN-I host
galaxies.  For a large sample of SLSNe-I, the average host-galaxy mass
increases with redshift; DES15E2mlf being the highest-redshift and
having one of the highest mass host galaxies of all SLSNe-I, continues this
trend.  We now explore some possible explanations.

First, most SLSNe are currently selected by the contrast in brightness
between the SN and its host galaxy.  Since SLSNe are $\gtrsim$2~mag
more luminous at peak than SNe~Ia, this strategy is effective at
finding particularly luminous SNe and lower luminosity SNe in
particularly low-luminosity host galaxies.  In practice, many SLSN
selection techniques prioritize SNe in faint or ``hostless'' galaxies 
\citep[i.e., galaxies below the detection limit of the images;][]{2015MNRAS.448.1206M}.  
As a result, there could be bias against SNe in detected galaxies.  
The luminosity of a galaxy at the
detection limit of a particular survey increases with redshift 
(i.e., only brighter galaxies can be detected at higher redshift).  
If SLSN host galaxies have a range of luminosities, a ``hostless''
selection criterion would remove SNe with the highest-mass host
galaxies at lower redshifts. In contrast, those highest-mass host
galaxies could be selected as hostless events at higher redshift. 
As a result, the average host-galaxy mass would increase with redshift.

However, \citet{2014ApJ...787..138L} found that this potential bias
was not significantly affecting the general trend that SLSNe are found
in lower-luminosity galaxies than other varieties of SNe.  
Even if the current SLSN sample could be biased to particularly
low-luminosity galaxies, there is likely a preference for SLSNe to
occur in such galaxies \citep{2014ApJ...787..138L,2016arXiv160408207P}.
Nonetheless, the fact that DES15E2mlf was originally selected as a likely
SN~Ia at low redshift, but has the higher-mass host galaxy of many
SLSN-I indicates that this bias certainly has some effect.  

While DES15E2mlf had a high-mass host galaxy, it could have
had a low-metallicity progenitor; the average metallicity evolves
significantly from $z \approx 2$ to $z = 0$.
To examine this possibility, we estimate the metallicity of the host
galaxy of DES15E2mlf using the mass-metallicity relationship
\citep[e.g.,][]{2004ApJ...613..898T}.  Since the relationship evolves
with redshift \citep{2013ApJ...771L..19Z, 2016MNRAS.456.2140M}, we
used the \citet{2013ApJ...771L..19Z} parameterization to determine the
mass-metallicity relation at $z = 1.861$.  With this relation and for
the measured \mstellar\ of the host of DES15E2mlf, we estimate the
host-galaxy has a gas-phase metallicity of $12 + \rm\log(O/H) =
8.72$\footnote{We used the \citet{2004ApJ...617..240K} metallicity
calibration as in \citet{2014ApJ...787..138L}.}, which is slightly above
the solar value \citep[8.69;][]{2001ApJ...556L..63A}. Despite the host galaxy's
relatively high mass, its metallicity is still $\sim$0.2\,dex lower
than that of low-$z$ field galaxies \citep[e.g.,][]{2008ApJ...681.1183K} at the same \mstellar.
A direct measurement would be critical to really constrain the host metallicity
of DES15E2mlf.

The detection of a SLSN in a relatively high-mass galaxy suggests that
other factors, such as progenitor metallicity, are more important in
creating such events.  At even higher redshifts, SLSNe may occur at a
much higher rate in higher-mass galaxies, perhaps providing
significant feedback and affecting galaxy formation and evolution.
SLSN searches that avoid selection based on host brightness could
improve our understanding of this process in high-mass galaxies and at
high redshift.

\section{Conclusions}
\label{sec:conclusion}
We present the discovery of DES15E2mlf, a SLSN, peaking at $M_{\rm AB}
= -22.3$\,mag, discovered in the DES-SN survey at $z = 1.861$. It is
the most distant SLSN confirmed to date. Spectroscopic comparisons
with other SLSNe indicate that DES15E2mlf is most similar to that of
hydrogen-poor SLSNe-I.

While the $riz$ light curves evolve similarly, the
$g$-band light curve (which corresponds to rest-frame 1700~\AA) peaks
$\sim$10 rest-frame days earlier while declining faster and more
dramatically than that of the other bands. We find excess flux in
the first detection of the $g$-band light curve. This could indicate
a potential precursor event similar to that found for many other
SLSNe. Compared to other SLSNe, we find the peak precursor 
luminosity is limited to $\rm M_{1700\AA} < -20.56$\,mag.
A blackbody temperature $T_{bb}\gtrsim$ 28,500\,K is determined
for the epoch of the first detection.

We obtained three spectra of DES15E2mlf, covering near-maximum phases
($-6$ to $+8$ rest-frame days).  The spectra have no strong absorption or
emission features between 2300 and 3500~\AA, but have two strong absorption features,
which were previously identified as \Feiii\ and
\ciii\ \citep{2013ApJ...779...98H,2016MNRAS.458.3455M}, at $\lambda \approx 1950$ and
2150\,\AA, respectively.  The wavelength of maximum absorption for
each feature shifts to the red with time.  Interpreting this as a
change in expansion velocity, the velocity decreases by $\sim$3000 --
6000\,km\,s$^{-1}$ for the time period covered by our spectra.

We detect strong narrow absorption features, which we interpret as
originating in the interstellar medium of the host galaxy.  Measuring the
equivalent widths of those features, we estimate the column density of
the interstellar medium along the line of sight. We find that
DES15E2mlf has strong \mgii\ absorption relative to the average GRB
absorbing system and the majority of quasar intervening systems.  This
shows the possibilities of using distant SLSNe to constrain the gas
content of the universe.

We examined the host properties of DES15E2mlf, finding that it has a
higher stellar mass than the typical SLSN-I host.  
We postulate that previous selection algorithms
that favored targeting SNe in faint galaxies could bias the
population statistics to lower-mass galaxies 
(at least for the high redshift SN samples) and that DES15E2mlf
avoided this potential bias since it was originally targeted as a
likely low-redshift SN~Ia.

Regardless of a potential selection bias, the host galaxy of
DES15E2mlf shows that SLSNe-I can occur in relatively normal galaxies 
(at least at high redshifts) and likely had a significantly
lower metallicity than local galaxies with the same mass.
Since DES15E2mlf is the highest-redshift SLSN-I known, the high-mass
of its host galaxy may simply be the result of decreasing metallicity
with redshift.  A direct metallicity measurement of the host galaxy
will result in a more robust conclusion.

Since SLSNe at higher redshift may occur more frequently in
higher-\mstellar\ galaxies, we suggest the future searches should
avoid selecting the candidates based on the brightness of host galaxy.
This is especially true for DES, which has become particularly adept
at discovering SLSNe at high redshift.

\section*{acknowledgments}
Y.-C.P.\ and R.J.F.\ are supported, in part, by NSF grant
AST-1518052.  Y.-C.P. thanks T.-W.~Chen for helpful discussions and
comments.  R.J.F.\ gratefully acknowledges support from the Alfred
P.\ Sloan Foundation and the David and Lucile Packard Foundation. 
L.G.\ was supported in part by the US National 
Science Foundation under Grant AST-1311862.
M.J.J.\ acknowledges support from the UK STFC [ST/N000919/1].
M.S.\ acknowledges support from EU/FP7-ERC grant number 615929.

Funding for the DES Projects has been provided by the U.S. Department of Energy, 
the U.S. National Science Foundation, the Ministry of Science and Education of Spain, 
the Science and Technology Facilities Council of the United Kingdom, the Higher 
Education Funding Council for England, the National Center for Supercomputing 
Applications at the University of Illinois at Urbana-Champaign, the Kavli 
Institute of Cosmological Physics at the University of Chicago, 
the Center for Cosmology and Astro-Particle Physics at the Ohio State University,
the Mitchell Institute for Fundamental Physics and Astronomy at Texas A\&M University, 
Financiadora de Estudos e Projetos, 
Funda{\c c}{\~a}o Carlos Chagas Filho de Amparo {\`a} Pesquisa do Estado do Rio de Janeiro, 
Conselho Nacional de Desenvolvimento Cient{\'i}fico e Tecnol{\'o}gico and 
the Minist{\'e}rio da Ci{\^e}ncia, Tecnologia e Inova{\c c}{\~a}o, 
the Deutsche Forschungsgemeinschaft and the Collaborating Institutions in the Dark Energy Survey. 

The Collaborating Institutions are Argonne National Laboratory, the University of 
California at Santa Cruz, the University of Cambridge, Centro de Investigaciones Energ{\'e}ticas, 
Medioambientales y Tecnol{\'o}gicas-Madrid, the University of Chicago, University College London, 
the DES-Brazil Consortium, the University of Edinburgh, 
the Eidgen{\"o}ssische Technische Hochschule (ETH) Z{\"u}rich, 
Fermi National Accelerator Laboratory, the University of Illinois at Urbana-Champaign, 
the Institut de Ci{\`e}ncies de l'Espai (IEEC/CSIC), 
the Institut de F{\'i}sica d'Altes Energies, Lawrence Berkeley National Laboratory, 
the Ludwig-Maximilians Universit{\"a}t M{\"u}nchen and the associated Excellence Cluster Universe, 
the University of Michigan, the National Optical Astronomy Observatory, the University of Nottingham, 
The Ohio State University, the University of Pennsylvania, the University of Portsmouth, 
SLAC National Accelerator Laboratory, Stanford University, the University of Sussex, Texas A\&M University, 
and the OzDES Membership Consortium.

The DES data management system is supported by the National Science Foundation under Grant Number AST-1138766.
The DES participants from Spanish institutions are partially supported by MINECO under grants 
AYA2012-39559, ESP2013-48274, FPA2013-47986, and Centro de Excelencia Severo Ochoa SEV-2012-0234.
Research leading to these results has received funding from the European Research Council under the 
European UnionÕs Seventh Framework Programme (FP7/2007-2013) including ERC grant agreements 240672, 291329, and 306478.
 
\bibliographystyle{mn2e}
\bibliography{des15e2mlf}

\section*{Affiliations}
\noindent$^{1}$Department of Astronomy and Astrophysics, University of California, Santa Cruz, CA 95064, USA\\
  $^{2}$Department of Physics \& Astronomy, University of Southampton, Southampton, Hampshire, SO17 1BJ, UK\\
  $^{3}$Pittsburgh Particle Physics, Astrophysics, and Cosmology Center (PITT PACC)\\ 
  $^{4}$Physics and Astronomy Department, University of Pittsburgh, Pittsburgh, PA 15260, USA\\
  $^{5}$Institute of Cosmology and Gravitation, Dennis Sciama Building, University of Portsmouth, Burnaby Road, Portsmouth PO1 3FX, UK\\
  $^{6}$Center for Mathematical Modelling, University of Chile, Beauchef 851, Santiago, Chile\\
  $^{7}$Millennium Institute of Astrophysics, Chile\\
  $^{8}$Astrophysics, The Denys Wilkinson Building, University of Oxford, Keble Road, Oxford, OX1 3RH, UK\\
  $^{9}$Department of Physics, University of the Western Cape, Bellville 7535, South Africa\\
  $^{10}$Kavli Institute for Cosmological Physics, University of Chicago, Chicago, IL 60637, USA\\
  $^{11}$Department of Astronomy and Astrophysics, University of Chicago, 5640 South Ellis Avenue, Chicago, IL 60637, USA\\
  $^{12}$Argonne National Laboratory, 9700 South Cass Avenue, Lemont, IL 60439, USA\\
  $^{13}$Australian Astronomical Observatory, North Ryde, NSW 2113, Australia\\
  $^{14}$School of Sciences, European University Cyprus, 6 Diogenis Str., Engomi, 1516 Nicosia, Cyprus\\
  $^{15}$Department of Physics and Astronomy, University of Pennsylvania 209 South 33rd Street, Philadelphia, PA 19104, USA\\
  $^{16}$Cerro Tololo Inter-American Observatory, National Optical Astronomy Observatory, Casilla 603, La Serena, Chile\\
  $^{17}$Department of Physics \& Astronomy, University College London, Gower Street, London, WC1E 6BT, UK\\
  $^{18}$Department of Physics and Electronics, Rhodes University, PO Box 94, Grahamstown, 6140, South Africa\\
  $^{19}$Fermi National Accelerator Laboratory, P. O. Box 500, Batavia, IL 60510, USA\\
  $^{20}$LSST, 933 North Cherry Avenue, Tucson, AZ 85721, USA\\
  $^{21}$CNRS, UMR 7095, Institut d'Astrophysique de Paris, F-75014, Paris, France\\
  $^{22}$Sorbonne Universit\'es, UPMC Univ Paris 06, UMR 7095, Institut d'Astrophysique de Paris, F-75014, Paris, France\\
  $^{23}$Kavli Institute for Particle Astrophysics \& Cosmology, P. O. Box 2450, Stanford University, Stanford, CA 94305, USA\\
  $^{24}$SLAC National Accelerator Laboratory, Menlo Park, CA 94025, USA\\
  $^{25}$Laborat\'orio Interinstitucional de e-Astronomia - LIneA, Rua Gal. Jos\'e Cristino 77, Rio de Janeiro, RJ - 20921-400, Brazil\\
  $^{26}$Observat\'orio Nacional, Rua Gal. Jos\'e Cristino 77, Rio de Janeiro, RJ - 20921-400, Brazil\\
  $^{27}$Department of Astronomy, University of Illinois, 1002 W. Green Street, Urbana, IL 61801, USA\\
  $^{28}$National Center for Supercomputing Applications, 1205 West Clark St., Urbana, IL 61801, USA\\
  $^{29}$Institut de Ci\`encies de l'Espai, IEEC-CSIC, Campus UAB, Carrer de Can Magrans, s/n,  08193 Bellaterra, Barcelona, Spain\\
  $^{30}$Institut de F\'{\i}sica d'Altes Energies (IFAE), The Barcelona Institute of Science and Technology, Campus UAB, 08193 Bellaterra Spain\\
  $^{31}$Department of Physics, IIT Hyderabad, Kandi, Telangana 502285, India\\
  $^{32}$Jet Propulsion Laboratory, California Institute of Technology, 4800 Oak Grove Dr., Pasadena, CA 91109, USA\\
  $^{33}$Instituto de Fisica Teorica UAM/CSIC, Universidad Autonoma de Madrid, 28049 Madrid, Spain\\
  $^{34}$Department of Astronomy, University of California, Berkeley,  501 Campbell Hall, Berkeley, CA 94720, USA\\
  $^{35}$Lawrence Berkeley National Laboratory, 1 Cyclotron Road, Berkeley, CA 94720, USA\\
  $^{36}$Astronomy Department, University of Washington, Box 351580, Seattle, WA 98195, USA\\
  $^{37}$Departamento de F\'{\i}sica Matem\'atica,  Instituto de F\'{\i}sica, Universidade de S\~ao Paulo,  CP 66318, CEP 05314-970, S\~ao Paulo, SP, Brazil\\
  $^{38}$George P. and Cynthia Woods Mitchell Institute for Fundamental Physics and Astronomy, Texas A\&M University, TX 77843,  USA\\
  $^{39}$Center for Cosmology and Astro-Particle Physics, The Ohio State University, Columbus, OH 43210, USA\\
  $^{40}$Department of Astronomy, The Ohio State University, Columbus, OH 43210, USA\\
  $^{41}$Instituci\'o Catalana de Recerca i Estudis Avan\c{c}ats, E-08010 Barcelona, Spain\\
  $^{42}$Department of Physics and Astronomy, Pevensey Building, University of Sussex, Brighton, BN1 9QH, UK\\
  $^{43}$Centro de Investigaciones Energ\'eticas, Medioambientales y Tecnol\'ogicas (CIEMAT), Madrid, Spain\\
  $^{44}$Department of Physics, University of Michigan, Ann Arbor, MI 48109, USA\\
  $^{45}$Universidade Federal do ABC, Centro de Ci\^encias Naturais e Humanas, Av. dos Estados, 5001, Santo Andr\'e, SP, Brazil, 09210-580\\
  $^{46}$Computer Science and Mathematics Division, Oak Ridge National Laboratory, Oak Ridge, TN 37831\\

\label{lastpage}

\end{document}